\documentclass[twocolumn]{aastex63}

\usepackage{amsmath}
\usepackage{amssymb}
\usepackage{amsthm}
\usepackage{natbib,aasdefs,url,bm}
\usepackage{graphicx}
\usepackage{color}
\usepackage{mathtools,  epsfig, lipsum}
\usepackage{subfigure}
\usepackage{textcomp}

\usepackage{multirow,booktabs,colortbl,tabularx}
\usepackage{enumitem}
\usepackage{booktabs}

\def\beq{\begin{equation}}
\def\eeq{\end{equation}}
\def\bey{\begin{eqnarray}}
\def\eey{\end{eqnarray}}

\shorttitle{GeV-to-TeV $\gamma$-rays from SS 433/W50}
\shortauthors{Fang et al.}

\begin{document}

\title{GeV-TeV Counterparts of SS 433/W50 from {\it Fermi}-LAT and HAWC Observations}

\author{Ke Fang} 
\affil{NHFP Einstein Fellow}
\affil{Kavli Institute for Particle Astrophysics and Cosmology (KIPAC), Stanford University, Stanford, CA 94305, USA}

\author{Eric Charles}  
\affil{SLAC National Accelerator Laboratory, 2575 Sand Hill Road, Menlo Park, CA 94025, USA}

\author{Roger D. Blandford} 
\affil{Kavli Institute for Particle Astrophysics and Cosmology (KIPAC), Stanford University, Stanford, CA 94305, USA}

\begin{abstract}
The extended jets of the microquasar SS 433 have been observed in optical, radio, X-ray, and recently very-high-energy (VHE) $\gamma$-rays by HAWC. The detection of HAWC $\gamma$-rays with energies as great as 25~TeV motivates searches for high-energy $\gamma$-ray counterparts in the {\it Fermi}-LAT data in the 100~MeV--300~GeV band. 
In this paper, we report on the first-ever joint analysis of {\it Fermi}-LAT and HAWC observations to study the spectrum and location of $\gamma$-ray emission from SS~433. 
Our analysis finds common emission sites of GeV-to-TeV $\gamma$-rays inside the eastern and western lobes of SS 433. The total flux above 1~GeV is $\sim 1\times10^{-10}\,\rm cm^{-2}\,s^{-1}$ in both lobes. The $\gamma$-ray spectrum in the eastern lobe is consistent with inverse-Compton emission by an electron population that is accelerated by jets. To explain both the GeV and TeV flux, the electrons need to have a soft intrinsic energy spectrum, or undergo a quick cooling process due to synchrotron radiation in a magnetized environment.  
\end{abstract}

\keywords{Gamma-ray sources, X-ray binary stars }

\section{Introduction}
SS 433 is a microquasar in the supernova remnant W50 (see \citealt{doi:10.1146/annurev.aa.22.090184.002451, Fabrika04} and references therein). It is likely composed of a $\sim$$20$~M$_\odot$ black hole orbiting a $\sim$$30\,M_\odot$ supergiant  companion with a 13.1~day period. The exotic system is located at a distance of $5.5$~kpc (\citet{2004ApJ...616L.159B}; see discussions about other distance measures in e.g., \citet{Marshall_2013}) and about $2^\circ$ below the Galactic plane. 
It produces two remarkable jets with kinetic power $L_{\rm kin}$$\sim$$10^{39}\,\rm erg\,s^{-1}$. The jets are heavily loaded with baryons and move at a speed of $0.26$~c while precessing with a period of $162$~days. The angle between jets and the axis is $\sim$$20^\circ$. 
Other periods are measured but the dynamics is poorly understood \citep{2001ApJ...561.1027E}.

\begin{figure*}[th]
\includegraphics[width= \linewidth] {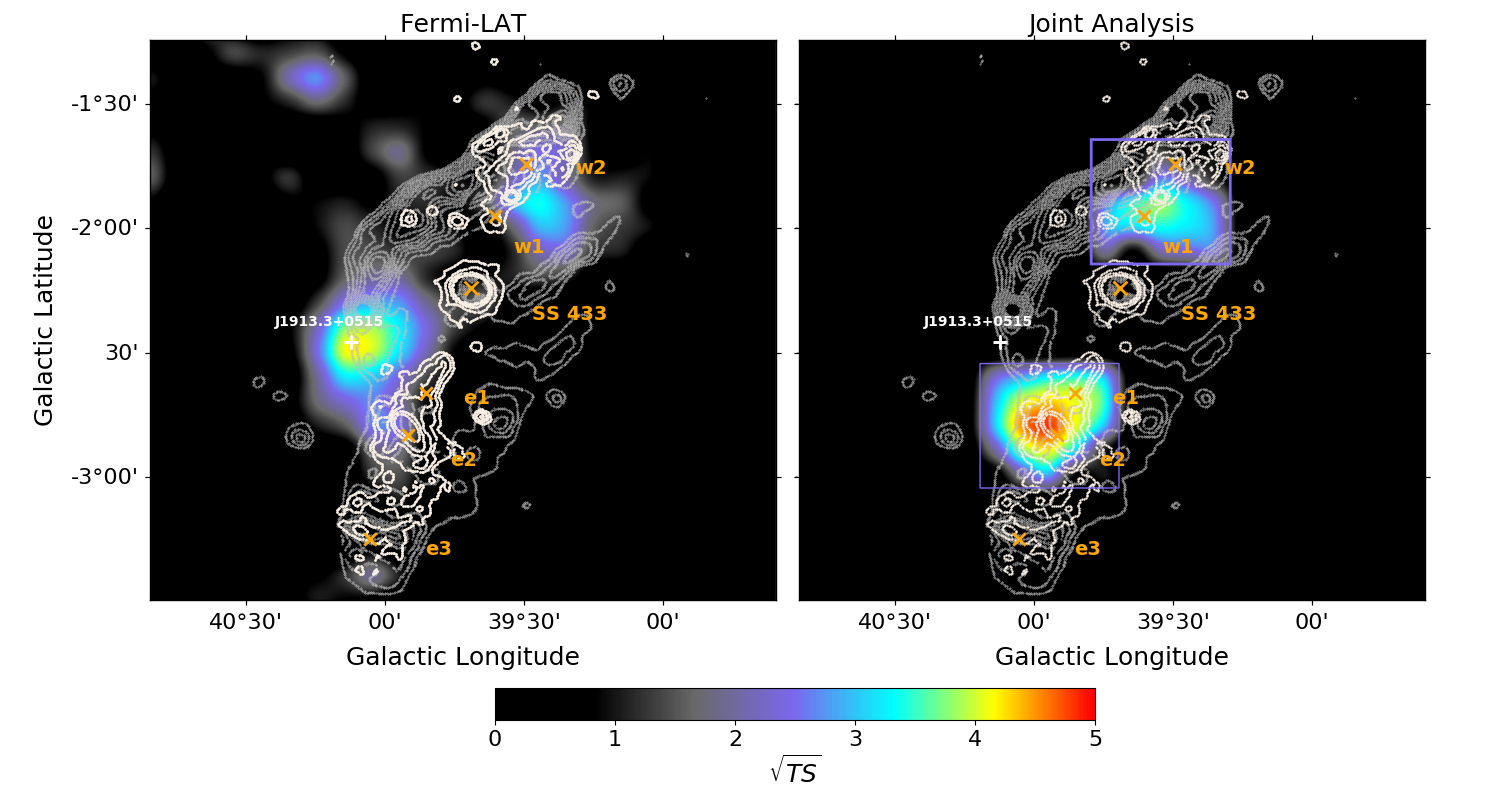}  
\caption{\label{fig:TSMap}
The SS 433/W50 region in the 10.5-year {\it Fermi}-LAT data between 100~MeV and 300~GeV (left) and from joint analysis of the {\it Fermi}-LAT data and the 1,017-day HAWC data (right) in Galactic coordinates. Left: The color scale indicates the statistical significance for a point source following an $E^{-2}$ spectrum as a function of position. The figure is a test statistic map after fitting $\gamma$-rays from known sources in the 4FGL catalog. Right: The background includes the 4FGL sources and J1913+0515 in the GeV band, and MGRO J1908+06 in the TeV band. The color scale indicates the improvement of the total likelihood of the ROI by a test point source that follows a log parabola spectrum in each $0.1^\circ\times 0.1^\circ$ grid inside the purple squares. The maps are smoothed by a Gaussian interpolation. The $\gamma$-ray hotspots revealed by joint analysis are inside the lobes and close to hard X-ray emission sites. 
For comparison, the locations of SS 433, the jet termination regions e1, e2, e3, w1 and w2 observed in the X-ray data are indicated as orange crosses. FL8Y J$1913.3+0515$ is marked by a white cross. The white and grey contours show the X-ray at $\sim0.9-2$~keV  \citep{1997ApJ...483..868S} and radio emission at 4.85~GHz \citep{1996ApJS..103..427G}. For a SS~433 distance of 5.5~kpc, 30’ corresponds to 50 pc.}
\end{figure*}

Extended X-ray jets are observed on the eastern and western sides (in Galactic coordinates) of SS 433 as shown by the white contours in Figure~\ref{fig:TSMap} \citep{1997ApJ...483..868S}. They interact with and distort the shell of the W50 nebula \citep{1983ApJ...273..688W, 1996ApJS..103..427G} which is shown by the grey contours. 
A set of emission regions, denoted as e1, e2, and e3 centered at $24'$, $35'$ and $60'$ east of SS 433, and w1 and w2 centered at $18'$ and $31'$ west of SS 433 have been investigated in detail  \citep{1997ApJ...483..868S}. A bright knot is seen in soft X-rays at e2 \citep{1997ApJ...483..868S, 2007A&A...463..611B}, and emission from e1, e2 and w1, w2 is observed in hard X-rays  \citep{1999ApJ...512..784S, 2005AdSpR..35.1062M}. The X-ray emission can be explained by synchrotron radiation of $\sim 100-200$ TeV electrons in a $\sim 10\,\mu$G magnetic field. 

Very-high-energy (VHE) $\gamma$-ray emission has recently been detected from the SS 433 lobes  by the High Altitude Water Cherenkov (HAWC) observatory \citep{hawcSS433}. In a dataset based on 1,017 days of measurements,  photons with energies of at least $25$~TeV are observed. 
The TeV hotspots are located close to e1, e2 and w1 with spatially unresolved emission profiles. The flux can be explained by the inverse-Compton emission of the same electron population whose synchrotron emission is observed by X-ray telescopes. On the other hand, 40--80~h observations with the Major Atmospheric Gamma Imaging Cherenkov telescopes (MAGIC) and High Energy Spectroscopic System (H.E.S.S.) reports no evidence of  $\gamma$-ray emission between a few hundred~GeV and a few TeV from the jet termination regions, nor from the central binary \citep{2018A&A...612A..14M}. A similar upper limit is reported by the Very Energetic Radiation Imaging Telescope Array System (VERITAS)  \citep{2017ICRC...35..713K}. 

Searches \citep{Bordas_2015, 2019ApJ...872...25X, 2019AA...626A.113S, 2019MNRAS.485.2970R} have been made for a GeV counterpart in the data observed by the {\it Fermi} Large Area Telescope (LAT) \citep{2009ApJ...697.1071A}.  
Analysis of LAT data in this region faces two complications. 
A point source FL8Y J1913.3+0515 from the preliminary LAT 8-year point source list\footnote{https://fermi.gsfc.nasa.gov/ssc/data/access/lat/fl8y/} (FL8Y) is tagged as possibly associated with W50. It is no longer a source in the 4FGL catalog due to different emission models \citep{4FGL}. Different conclusions about detection of SS 433 have been reached depending on whether this source is included in the background model. In addition, analysis of the emission profile and spectrum of the SS 433 region is heavily impacted by the nearby pulsar PSR J$1907+0602$ ($\rm 4FGL\,J1907.9+0602$), which is not suppressed via selection on rotational phase in the above-mentioned works. Due to these difficulties, whether the jets of SS~433 shine at GeV energies is unknown.

Since the source is marginally significant in $\gamma$-rays yet close to the bright Galactic plane, it is difficult to study solely with the GeV or the TeV measurements. Here we jointly analyze a region-of-interest (ROI) surrounding SS 433 observed by LAT and  HAWC. A simultaneous fit to the 100~MeV to 100~TeV data directly addresses the question whether $\gamma$-ray emission over six decades can be produced by a common cosmic-ray population inside the SS 433 lobes. 
We explain the methods in Section~\ref{sec:dataAnalysis}, present the results of the LAT analysis in Section~\ref{subSec:fermiResults} and that of the joint analysis in Sections~\ref{subSec:J1913} and \ref{subSec:jointAnalysisResults}. We discuss immediate implications of this analysis in Section~\ref{sec:discussion}. 

\section{Methods}\label{sec:dataAnalysis}

Our analysis uses 10.5 years of {\it Fermi}-LAT data and 1,017 days HAWC data. Details of the LAT and HAWC analyses, as well as background sources in each band are presented in Appendices~\ref{appendix:Fermi-LAT} and \ref{appendix:HAWC}. The setup of a joint analysis of the LAT and HAWC data based on the 3ML framework  \citep{3ML}
is presented in Appendix~\ref{appendix:jointAnalysis}. 
Here we describe the procedure for the joint analysis.

We first build a source model to describe the broadband $\gamma$-ray emission of SS 433. Three types of models are considered. 
\def\theenumi{\Roman{enumi}}

 \begin {enumerate}  
 \item $\gamma$-rays follow a power-law spectrum, $dN/dE_\gamma=K_\gamma\,\left(E_\gamma/E_{\gamma, \rm piv}\right)^{-\alpha_\gamma}$. 
  \item $\gamma$-rays follow a LogParabola spectrum, $dN/dE_\gamma=K_\gamma\,\left(E_\gamma/E_{\gamma, \rm piv}\right)^{-\alpha_\gamma - \beta_\gamma\,\log(E_\gamma/E_{\gamma,\rm piv})}$. 
  \item Electrons are injected with a rate $d\dot{N}/dE_e = Q_e\,\gamma_e^{-\alpha_e}\,\exp\left({-E_e / E_{e,\rm max}}\right)$, where $\gamma_e=E_e/m_e\,c^2$ is the Lorentz factor of an electron. They up-scatter the cosmic microwave background (CMB) and infrared photons in W50 to $\gamma$-rays through the inverse-Compton process, and produce synchrotron emission in a magnetic field B. 
 \end {enumerate}

Models I and II are simple descriptions of $\gamma$-ray spectral shapes. Model III is physically motivated. The cooling time of VHE electrons, $t_{e, \rm cool} = 2.5\,(B/10\,\mu\rm G)^{-2}\,(\gamma_e/10^8)^{-1}$~kyr, is much less than the source age, $t_{\rm age} \sim 30\,\rm kyr$ \citep{Fabrika04}. In order to take into account the effects of cooling, we solve a transport equation for each set of parameter values. Details about the cooled electron spectrum can be found in Appendix~\ref{appendix:cooling}. 
Once a steady-state cosmic-ray spectrum is obtained, the $\gamma$-ray flux is calculated using the radiative functions of the {\it naima} package \citep{naima, 2014ApJ...783..100K}.

The models are then converted into data space and compared to observation through the joint analysis framework (see Appendix~\ref{appendix:jointAnalysis}). Since the GeV and TeV observations are carried out independently, a total log likelihood is evaluated by summing the log likelihoods from the GeV and TeV analyses. The total likelihood is then maximized by adjusting model parameters to obtain the best-fit source model. Finally, the likelihood test statistic (TS) of a target source is computed as twice the difference of the log likelihoods of the data given the models with and without the source.

\section{Results}\label{sec:results}
\newcolumntype{C}[1]{>{\centering\arraybackslash}p{#1}}

\begin{table*}[htbp]
\caption{Fit results} \label{table:models}
\centering
   \begin{tabular}{C{2.3cm} C{2.8cm} C{2cm}*{4}{C{1.5cm}} }
\toprule
    Source &
    Position &
    \multicolumn{2}{p{4cm}}{\centering{TS (Individual)}} &  
    \multicolumn{1}{p{1.2cm}}{\centering{ Model*}}& 
     \multicolumn{2}{p{3cm}}{\centering{Significance}} \\
  \cmidrule(lr){3-4}    \cmidrule(lr){6-7}
   &  (R.A., Dec. in degree)
      &
    LAT & HAWC  & 
  & Individual   & Total \\
  \midrule
 eastern hotspot   &   (288.56,  4.95)    &   1.9  &   21.6   & I   &  $4.2\,\sigma$   &  $5.5\,\sigma$   \\        
 western hotspot   &   (287.58,  5.01)   &  8.9 &   12.1  & I & $3.9\,\sigma$  &     \\  
  \addlinespace[-2ex] 
\midrule
 eastern hotspot   &   (288.56,  4.95)    &   4.3  &   21.7   & II   &  $4.4\,\sigma$   &  $5.4\,\sigma$   \\        
 western hotspot   &   (287.58,  5.01)  &   4.6 &  12.4  & II  & $3.4\,\sigma$  &     \\  
   \addlinespace[-2ex] 
 \midrule
  eastern hotspot   &   (288.56,  4.95)    &  3.3  &   19.9  & III   &     $4.1\,\sigma$   &  $5.0\,\sigma$   \\        
 western hotspot   &   (287.58,  5.01)   &  5.2 &   10.8   & III & $3.3\,\sigma$  &     \\         
   \addlinespace[-2ex]                    
\bottomrule
\end{tabular}
   \\[10pt]
{* For particular models, certain parameters are held constant. They include: Model I, $E_{\rm piv} =  875.753 \, {\rm MeV}$, $\alpha_{\gamma, W}=2.2$, $\alpha_{\gamma, E}=2.1$;
Model II, $\alpha_\gamma=1.8$, $\beta_\gamma=0.05$, $E_{\gamma,\rm piv}=60$~GeV;
Model III, $\alpha_e=1.9$, $B=20\,\mu$G, and $E_{e,\rm max}=1\,\rm PeV$. RA and Dec are for epoch J2000. See text for additional details. 
 }

\end{table*}

\subsection{LAT analysis results} \label{subSec:fermiResults}

Here we present results from the LAT-only analysis. The method is detailed in Appendix~\ref{appendix:Fermi-LAT}. The main difference between our analysis and previous works is that we use a dataset for which the PSR J1907+0602 is gated off, that is, the arrival times of photons are phase folded, and the photons that arrive during the pulsar's pulse peak are removed (see \citealp{JianLi} for details). 
Throughout the work we use the 4FGL catalog and the corresponding diffuse emission models to model background sources.  
 
The significance of the residual $\gamma$-ray excess from the SS 433/W50 region in the LAT data between 100~MeV and 300~GeV is shown in the left panel of Figure~\ref{fig:TSMap}. The most statistically significant excess is near the location of FL8Y J1913.3+0515. We call this excess J1913+0515 to differentiate it from FL8Y J1913.3+0515. J1913+0515 is at the boundary of W50, well outside the extended X-ray jets. When describing the SED with a power-law function, we obtain $K_\gamma= 1.5\times10^{-12}\,\rm MeV^{-1}\,cm^{-2}\,s^{-1}$, $E_{\gamma,\rm piv} = 0.9\,\rm GeV$ and $\alpha_\gamma =2.4$. The best-fit location is very close but slightly different from the location listed in the FL8Y catalog. The test statistic of the source is $\rm {TS} =32.8$ using a regular likelihood and $\rm {TS}=25.7$ using a weighted likelihood \citep{4FGL} that takes into account estimated systematic uncertainties in the diffuse emission models. As the results obtained by the two methods are similar, we use a regular (i.e., unweighted) likelihood in the rest of our analyses.

A sub-threshold (TS $< 25$) excess is evident at the northeastern side of J1913+0515. Because it is spatially close to the TeV excess in the eastern lobe, we refer to it as the ``eastern hotspot". It is not significant in the LAT data and has ${\rm TS}= 5.0$ when J1913+0515 is included in the background model. The excess is due to several high-energy photons at $\sim 20- 50$~GeV, as shown by the SED in the top panel of Figure~\ref{fig:PL_e}.  

In the western lobe, a sub-threshold excess is found between w1 and w2 (which we refer to as the ``western hotspot" below). The excess region partially overlaps with the X-ray jets and touches the boundary of W50.  We find a TS of 16.1 for the western hotspot when adding it to the baseline model.  Its spectrum can be described by a power law of index $2.3$ as shown in the bottom panel of Figure~\ref{fig:PL_e}.

When including the potential sources in the baseline model simultaneously, we obtain ${\rm TS}\sim 5$ and ${\rm TS}\sim 10$ for the eastern and western hotspots, respectively. 
The fit results are summarized in Table~\ref{table:Fermi}.  Neither of the ``hotspots" is statistically significant in the LAT observations but they are evident in the joint analysis as will be shown in Section~\ref{subSec:jointAnalysisResults}.

\subsection{J1913+0515 and the TeV emission}\label{subSec:J1913}

To investigate whether J1913+0515 and the TeV emission  in the eastern lobe share a common origin, we test two ways of combining the GeV and TeV hotspots. 

First, we replace J1913+0515 and the TeV excess with a single source centered between them and assume that it has a power-law spectrum. 
The joint fit has six free parameters in total, including spectral index, flux normalization, extension of  MGRO J1908+06, and flux normalization and location (RA, Dec) of the test source. 
Due to the low statistics, it is difficult to fit the spectral index and the flux normalization of the test source simultaneously. We thus fix the index as $\alpha_\gamma=2.2$ and vary only the prefactor. 
Following \citet{wilks1938} theorem and \citet{chernoff1954}, we calculate the probability of the TS using a chi-square distribution with three degrees of freedom, which is the difference in dimensionality of the models when including and excluding the test source.  We then evaluate the corresponding number of standard deviations for this confidence level for a Gaussian distribution.

We find ${\rm TS} = 32.1$ for the test source, which includes $16.9$ from a comparison with the LAT data,  and $15.2$  from a comparison with the HAWC data.  The TS corresponds to $5.0\,\sigma$ standard deviation.

Alternatively, we assume that the sources share a spectrum but differ in emission sites. 
The fit results in ${\rm TS}=30.3$ from GeV data and ${\rm TS}=23.8$ from TeV data. The total significance increases to $6.4\,\sigma$, despite the two extra degrees of freedom due to the additional emission site. In general, we find that the LAT TS of the common source increases and the HAWC TS decreases when the test source is moved toward J1913+0515, and the trend is reversed when the test source is moved toward the VHE hotspot. Such a trend, along with the considerable difference in the statistical significances of the models  with one and two source locations, suggest that J1913+0515 is unlikely to be a counterpart of the TeV hotspot in the eastern lobe.

\subsection{Joint analysis results}\label{subSec:jointAnalysisResults}

 \begin{figure}
\includegraphics[width= \linewidth] {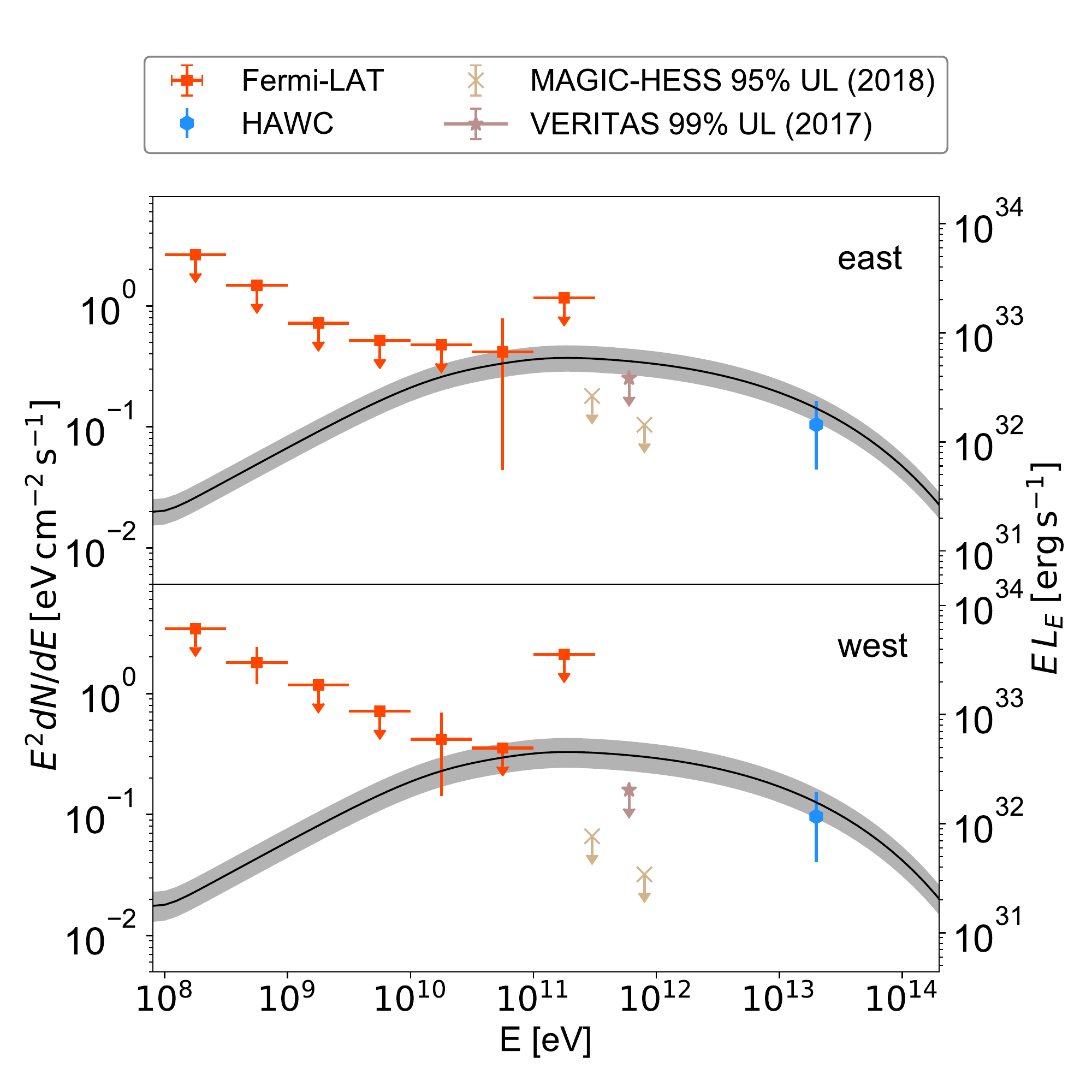}  
\caption{\label{fig:PL_e}
The best-fit $\gamma$-ray spectra in the eastern and western lobes obtained by joint analysis assuming that $\gamma$-rays are produced by an electron population (Model III). The parameters and TS of the model are listed in Table~\ref{table:models}. The grey shaded area indicates the $68\%$ statistical uncertainty from a fit that varies the normalization. For comparison, we show the SED from the LAT-only analysis (Section~\ref{subSec:fermiResults}, red markers), HAWC-only analysis (Appendix~\ref{appendix:HAWC}, \citealt{hawcSS433}; blue markers), upper limits on $\gamma$-rays from nearby regions by VERITAS \citep{2017ICRC...35..713K} and HESS \citep{2018A&A...612A..14M} (grey markers). For the LAT data points, 95\% upper limits are shown when TS~$<4$, otherwise $1\,\sigma$ error bars are shown. Since IACT limits are converted from integral limits, they do not have horizontal error bars. We find that the $\gamma$-ray emission in the eastern lobe can be explained as the inverse-Compton emission by a cooled electron population. 
}
\end{figure}

Motivated by results from the last section, we perform a joint analysis of LAT and HAWC data with J1913+0515 added to the background. The parameters and TS of the models are summarized in Table~\ref{table:models}. For rectangular areas with $\Delta\,l = 0.5^\circ$ and $\Delta\,b = 0.5^\circ$ that cover the e1, e2 and w1, w2 regions, we compute the TS of a test source at every $0.1^\circ\times 0.1^\circ$ grid point assuming a log parabola spectrum (Model II). The scanned regions are enclosed by the purple squares in Figure~\ref{fig:TSMap}.    A log parabola spectrum is chosen because it describes the LAT SED and  the HAWC flux better than a power-law spectrum. We take Model II with $\alpha_\gamma=1.8$, $\beta_\gamma=0.05$, $E_{\gamma,\rm piv}=60$~GeV but have verified that alternative log parabola shapes (for example with $\alpha_\gamma=1.7$, $\beta_\gamma=0.05$, $E_{\gamma,\rm piv}=5$~GeV) leads to similar results. We leave the normalization $K_\gamma$ as a free parameter. For background sources, we free the normalization and index of J1913+0515, the normalization and extension of MGRO J1908+06, and fix parameters of the rest of sources in the ROI. The map of TS for the best source positions considered is shown in the right panel of Figure~\ref{fig:TSMap}. We find that when including the TeV data, the eastern hotspot becomes significant, and can be resolved from J1913+0515.

To directly check whether the GeV-to-TeV emission can be explained as inverse-Compton emission of the same electron population, we perform a joint fit with the electron model (Model III). We fix the parameters in both lobes as $\alpha_e=1.9$, $B=20\,\mu$G, and $E_{e,\rm max} = 1\,\rm PeV$ and free the normalizations of the electron spectra. These parameters and their values are motivated by the fit to the broadband multi-wavelength data in \citet{hawcSS433}.
We do not scan the parameter space for these parameters, but note that the electron energy needs to be higher than 150~TeV to produce the measured 20~TeV photons. In general higher $E_{e,\rm max}$ leads to better fits. 
The best-fit model has a TS of 40 when fitting both lobes simultaneously. With 6 free parameters including the two normalizations and the coordinates of the two hotspots, the TS corresponds to a significance of $5\,\sigma$ for a two-sided Gaussian distribution. The fit results using all three models are listed in Table~\ref{table:models}. They are all significant, suggesting that the GeV-to-TeV $\gamma$-rays can be explained by common sources inside the SS 433 lobes. 

The SED is shown in Figure~\ref{fig:PL_e}. For comparison, we also show the SED obtained from the LAT-only analysis, the upper limits (UL) on nearby $\gamma$-ray emission by imaging air Cherenkov telescopes (IACT),  
and the flux at the pivot energy $E_{\rm piv}=20\,\rm TeV$ from the HAWC-only analysis. 
 We find that the $\gamma$-ray flux and the cosmic-ray injection rates of the east and the west hotspots are very similar. 
 To explain both the GeV and TeV flux, a  soft electron spectrum $dN/dE\sim E^{-3}$ is needed. This can be achieved by a relatively inefficient acceleration~\citep{BLANDFORD19871} or by cooling of electrons as suggested by \citet{hawcSS433}. The flux at GeV energies is higher than that predicted by \citet{hawcSS433}. This suggests that a far-infrared background needs to be present (whose energy density is discussed in Appendix~\ref{appendix:cooling}).

The best-fit models predict a sub-TeV $\gamma$-ray flux that is higher than the upper limits from IACTs. The upper limits are based on observations of e2, w2 (H.E.S.S.) and w1 (VERITAS) with small angular extents defined by X-ray observations, and so they do not necessarily apply to the actual source locations in these models.

The western source is less significant in all cases, which could be due to confusion by Galactic diffuse emission and with MGRO J1908+06. The location of the $\gamma$-ray emission site in the western lobe is less clear. Like the eastern side, the localized GeV emission could be a combination of emission inside the lobe and at the boundary of W50, though more statistics is needed to verify this scenario.

\section{Discussion}\label{sec:discussion}
Because of its proximity and exotic structure, SS~433 has been one of the most observed Galactic high-energy sources for over 40 years. Nonetheless, no consensus has been reached about what happens inside the SS~433/W50 complex. The detection of multi-tens of TeV photons from the object confirms  the existence of particles at extreme energies, but deepens the question why lower-energy $\gamma$-rays have not been observed. By jointly analyzing an ROI measured by both the {\it Fermi}-LAT and the HAWC Observatory, we find common sites of GeV and TeV $\gamma$-ray emission inside the SS 433 lobes. The spectral energy distribution is consistent with inverse-Compton emission of an electron population accelerated by the jets but quickly cooled due to synchrotron radiation in a magnetized environment. We use a dataset that suppresses emission by a nearby pulsar that highly impacts previous analyses. Our joint analysis concludes that the GeV point source J1913+0515 located at the boundary of W50 is unlikely a counterpart to the TeV emission. This addresses the dilemma encountered by \citet{2019ApJ...872...25X, 2019AA...626A.113S, 2019MNRAS.485.2970R}.

This is the first joint-ROI analysis across $\gamma$-ray observatories to our knowledge. Using a framework built on individual data analysis toolkits from $\gamma$-ray observatories, we have shown that such an approach is feasible. The joint analysis is designed to study shared properties of sources of $\gamma$-rays over a very broad spectrum. It maximizes the usage of data including sub-threshold information, and is more powerful than simply combining results from each experiment. 
Future data from HAWC, especially with refined angular resolutions \citep{2019arXiv190512518H} will help improve the understanding of $\gamma$-ray emission from the western lobe. Future observations by IACTs, as well as by X-ray and radio telescopes, at the revised source locations will help further constrain the emission models.

The GeV-TeV spectrum can be explained as inverse-Compton scattering by X-ray synchrotron-emitting $\sim100\,{\rm TeV}$ electrons. Three scenarios can be entertained to account for the acceleration.  The first is to invoke direct, diffusive shock acceleration of the electrons at the termination shocks of the precessing jets launched by the accretion disk. If the post-shock field strength is $\sim10\,\mu{\rm G}$ then acceleration to these energies is possible, though only a small electron power $\sim10^{34}\,{\rm erg\,s}^{-1}$ is needed to account for the $\gamma$-rays. Secondly, $\sim5\,{\rm PeV}$ protons may also be shock-accelerated. Their primary radiative loss could be due to Bethe-Heitler pair production on $\sim2\,{\rm eV}$ optical photons from SS 433 with a cross-section of a few millibarn. Maximum electron or positron energies $\sim100\,{\rm TeV}$ are just possible. However, in order to account for the $\gamma$-ray power, a proton power $\sim10^{38}\,{\rm erg\,s}^{-1}$ is necessary. The third possibility is that a hitherto unobserved, ultra-relativistic, electromagnetic jet is formed by the spinning black hole.
Such a jet can create an EMF $\sim100\,(L_{\rm jet}/10^{39}\,{\rm erg\,s}^{-1})^{1/2}\,{\rm PV}$ 
that suffices to accelerate the emitting particles. Finally, if the gas density in the lobes is high ($\gtrsim 1\,\rm cm^{-3}$), pion production can also make a contribution to the $\gamma$-ray flux. High-energy neutrinos would be produced simultaneously and could be measured by IceCube \citep{2018arXiv181107979I}. 

Future multi-messenger observations of the $\gamma$-ray emission regions have the potential to discriminate between these scenarios.

\bigskip
The \textit{Fermi} LAT Collaboration acknowledges generous ongoing support
from a number of agencies and institutes that have supported both the
development and the operation of the LAT as well as scientific data analysis.
These include the National Aeronautics and Space Administration and the
Department of Energy in the United States, the Commissariat \`a l'Energie Atomique
and the Centre National de la Recherche Scientifique / Institut National de Physique
Nucl\'eaire et de Physique des Particules in France, the Agenzia Spaziale Italiana
and the Istituto Nazionale di Fisica Nucleare in Italy, the Ministry of Education,
Culture, Sports, Science and Technology (MEXT), High Energy Accelerator Research
Organization (KEK) and Japan Aerospace Exploration Agency (JAXA) in Japan, and
the K.~A.~Wallenberg Foundation, the Swedish Research Council and the
Swedish National Space Board in Sweden.
 
Additional support for science analysis during the operations phase is gratefully
acknowledged from the Istituto Nazionale di Astrofisica in Italy and the Centre
National d'\'Etudes Spatiales in France. This work performed in part under DOE
Contract DE-AC02-76SF00515.

We thank Henrike Fleischhack, Colas Rivi\'ere, and Giacomo Vianello for their help with the usage of the {\it 3ML} and {\it HAWC-HAL} packages. We thank Jian Li and Matthew Kerr for their help with performing the pulsar gating in the LAT data analysis.

\bibliography{GeV_TeV_SS433}

\appendix
\restartappendixnumbering

\section{{\it Fermi}-LAT Analysis}\label{appendix:Fermi-LAT}


\begin{table}[htbp]
\caption{Significance of the candidate sources in the LAT data} \label{table:Fermi}
\centering
   \begin{tabular}{C{3cm} C{4cm} C{3cm} C{3cm}  C{1.2cm}}
\toprule
   Note &  Source & Position (RA, Dec in degree, J2000) & 1$\sigma$ uncertainty (in degree)    & TS \\
\midrule
Fit individually &  J1913+0515    &   (288.30,  5.24) & 0.06   &  32.8       \\        
With J1913  &  eastern hotspot   &   (288.56,  4.95)  & 0.27 & 5.0    \\        
 Fit individually   &  western hotspot    &   (288.53,  4.93) & 0.13  &  16.1    \\        
  \midrule                        
Fit two sources simutaneously &  J1913+0515    &   (288.31,  5.24)  & 0.06 &  28.4       \\       
 \addlinespace[-2ex] 
 & western hotspot   &  (287.58,  5.01)   & 0.16 &  9.7    \\
  \addlinespace[-2ex] 
  \midrule  
Fit three sources simultaneously &  J1913+0515    &   (288.31,  5.24) & 0.06  &  26.1      \\   
 \addlinespace[-2ex]    
 & eastern hotspot   &  (288.56,  4.95) & 0.33   &   5.0      \\  
  \addlinespace[-2ex]
 & western hotspot  &  (287.58,  5.01) & 0.17  &   9.6    \\ 
  \addlinespace[-2ex]
\bottomrule
\end{tabular}
\end{table}

We analyze $10.5$ years of Pass 8 data\footnote{https://fermi.gsfc.nasa.gov/ssc/data/analysis/documentation/Cicerone/Cicerone\_Data/LAT\_DP.html} taken between 2008-08-04 15:43:36 UTC and 2019-01-28 00:00:00 UTC using version 0.17.4 of {\it fermipy}\footnote{https://github.com/fermiPy/fermiPy} and version ScienceTools-11-04-00 of {\it Fermitools}\footnote{https://github.com/fermi-lat/ScienceTools}. We define the ROI as the $15^\circ\times 15^\circ$ region in Galactic coordinates centered at SS 433 ($l=39.69,\,  b=-2.24$). $\gamma$-ray events with energies between $100$~MeV and $300$~GeV are selected. Other event selection criteria include a $90^\circ$ zenith cut and a filter expression of ``DATA\_QUAL $>0$ \&\& LAT\_CONFIG $==$ 1" which are  standard quality criteria recommended by the {\it Fermi} Science Support Center. We use the {\it P8R2\_SOURCE} event selection, and the corresponding {\it P8R2\_SOURCE\_V6} LAT instrument response functions. Unlike previous works analyzing the SS 433 region \citep{2019ApJ...872...25X, 2019AA...626A.113S, 2019MNRAS.485.2970R}, here we use the latest LAT 8-year Point Source Catalog   \citep{4FGL}, together with the corresponding Galactic diffuse model {gll\_iem\_v07.fits} and the isotropic diffuse model. The 4FGL catalog and the updated diffuse emission model turn out to considerably impact the analysis of 100--300~MeV photons in this region, comparing to the FL8Y catalog. 

As the nearby pulsar PSR J1907+0602 is very bright in the GeV band, we follow the method from \cite{JianLi} to suppress the pulsar emission. The same pulsar ephemeris is adopted in pulsar gating which amounts to 44\% of the observing time. The exposure is scaled accordingly.

There are 33 4FGL sources within $5^\circ$ of SS 433 and 61 4FGL sources within $8^\circ$. The baseline ROI analysis is performed using the {\it fermipy.job} sub-package {\it fermipy-analyze-roi}. The optimized model is referred to as ``baseline model".  

A significance map of the residual $\gamma$-ray excess is shown in the left panel of Figure~\ref{fig:TSMap}. The color scale corresponds to the square root of the TS when there is a new point source at a given location, in addition to known sources from the 4FGL catalog, the Galactic diffuse emission and the isotropic diffuse emission. The test point source is assumed to have an $E^{-2}$ spectrum and the TS is evaluated for each location on a grid with $0.1^\circ\times 0.1^\circ$ spacing.  

After setting up the baseline model, we add J1913+0515, the eastern and western hotspots to the background model and refit the new model to the data. The new model is re-fit to the data using {\it GTAnalysis.optimize}, which fits sources in the order of their fluxes. The fit returns TS values of 26.1, 5.0 and 9.6 for the three candidate sources, respectively. We also tested an alternative fitting method, where we fixed the parameters of background sources to their best-fit values, and vary only the normalization of new sources using {\it GTAnalysis.fit}. This approach returned TS values of  28.0, 4.9 and 10.4 for the three. Since the difference of the results from the two fitting methods is minor while the latter is much more efficient in computation time, we use the second approach to calculate the LAT likelihood in a joint analysis.  

Although MGRO J1908+06 is one of the brightest TeV sources, an extended source at its location is not significant in the LAT data. We thus do not include it in our background model. 

The data points for the spectral energy distribution (SED) were obtained by binning the spectrum with 2 bins per decade in energy and performing a likelihood analysis in each energy bin.

\section{HAWC Analysis}\label{appendix:HAWC}
In the TeV band, we analyze the public data\footnote{https://data.hawc-observatory.org/datasets/ss433\_2018/index.php} from the High Altitude Water Cherenkov (HAWC) Observatory    \citep{hawcSS433}. The dataset contains 1,017 days of $\gamma$-ray events collected between 26 November 2014 and 20 December 2017.  The reconstruction of the arrival direction of primary $\gamma$-rays is based on the relative arrival times of photoelectron hits detected by the photomultipliers inside the water Cherenkov detectors. (This kind of reconstruction is referred as the nhit method.) Angular resolution from the nhit analysis ranges from around $1^\circ$ below 1~TeV to $<0.2^\circ$ above 10~TeV.

We adopt the same ROI as in \citet{hawcSS433}, which is defined to be a semicircular region with a radius of $2.5^\circ$ centered on the position of MGRO J1908+06 (as shown in Extended Data Fig 1 of \citealp{hawcSS433}). By masking the sources close to the Galactic plane, the contamination from the Galactic diffuse emission is significantly reduced.  

Three sources remain in the ROI: MGRO J1908+06, the eastern and the western hotspots in the SS~433 lobes. Following \citet{hawcSS433}, we use the electron diffusion model to describe the spatial morphology of MGRO J1908+06. Other spatial models with Gaussian and power-law radial profiles lead to similar results.   
 
Unlike the analysis in \citet{hawcSS433}, which is based on the HAWC analysis framework AERIE, here we redo the analysis using the HAWC Accelerated Likelihood (HAL) framework. HAL provides faster convolution with the detector response functions which is needed by the joint analysis in our work. We have confirmed that the two analysis frameworks lead to results that are consistent at the 1\% level.

\section{Joint Analysis}\label{appendix:jointAnalysis}

\begin{figure}[htbp]
\centering
\includegraphics[width= 0.7\linewidth] {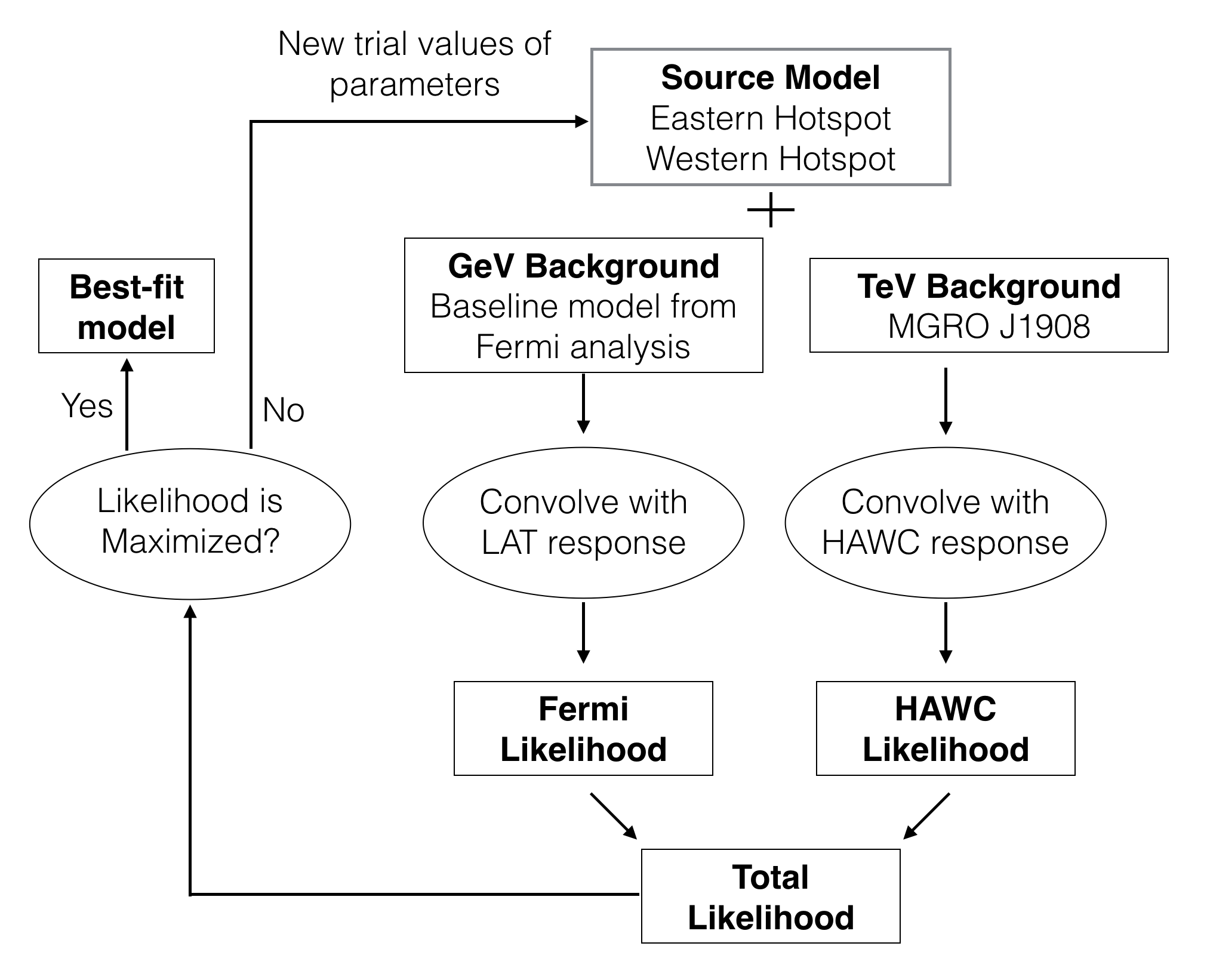}  
\caption{\label{fig:flowChart}
Diagram of the work flow of a joint analysis of the {\it Fermi}-LAT and HAWC data around SS 433. The source model describes the $\gamma$-ray emission of the potential sources and may depend on the parent particle types, injection spectra, source locations, and magnetic field strength. The background model is composed of dozens of 4FGL sources, the diffuse emission models in the GeV band, and MGRO J1908+06 in the TeV band. The models are passed to analysis pipelines for the two data sets, converted to data space and compared to data separately. The total likelihood is maximized to obtain the best-fit model. 
}
\end{figure}

The work flow of a joint analysis is diagrammed in Figure~\ref{fig:flowChart}. 
The joint analysis is implemented in the Multi-Mission Maximum Likelihood framework (3ML) \citep{3ML}\footnote{https://github.com/threeML/threeML}. 3ML is a data analysis architecture that converts emission models for an ROI into data spaces for specified instrument(s), and compares the model predictions to the corresponding data based on the likelihood formalism. The {\it Fermi}-LAT module of the package provides a wrapper of {\it fermipy} \citep{fermipy} and the {\it Fermi Science Tools}\footnote{https://fermi.gsfc.nasa.gov/ssc/data/analysis/software/}. The HAWC module links to the HAWC analysis tools including the {\it Accelerated Likelihood} (HAL) framework\footnote{https://github.com/threeML/hawc\_hal}. 

Since a ROI analysis is not implemented in 3ML, we use {\it fermipy} to perform a ``baseline" analysis externally (see Appendix~\ref{appendix:Fermi-LAT}) and use that as a starting point for 3ML analysis.
Meanwhile, the source model and a model of MGRO J1908+06 are passed to the HAWC plugin (see Appendix~\ref{appendix:HAWC}). 
In this way the contribution of background sources is taken into account properly.

\section{Radiative cooling of electrons}\label{appendix:cooling}
The cooling of relativistic electrons in the lobes of SS 433 can be described by a transport equation 
\beq\label{eqn:transport}
\frac{\partial N_e}{\partial t} + \frac{\partial}{\partial \gamma_e}\left[\dot{\gamma_e}\,N_e(\gamma_e, t)\right] = Q_e(\gamma_e, t), 
\eeq
where $\dot{\gamma}_e = -{4}/{3}\,\gamma_e^2\, c\, \sigma_T\,(u_B + u_{\rm \gamma}) / (m_e\,c^2)\equiv -\nu\,\gamma_e^2$ is the energy loss rate due to inverse-Compton and synchrotron emission. $\sigma_T$ is the Thomson cross section. $N_e$ and $Q_e$ are the spectrum and injection rate of electrons, respectively. $u_B = B^2 / (8\pi)$ and $u_{\rm CMB} = 0.26\,\rm eV\,cm^{-3}$ are the energy density of magnetic field and the CMB. We also adopt a far-infrared (FIR) background at $20$~K with $u_{\rm FIR} = 0.3 \,\rm eV\,cm^{-3}$ motivated by the dust emission in the solar neighborhood \citep{2016PhRvD..94f3009V}. Background photons with higher energies are not important due to the Klein-Nishina effect. Due to its much lower energy density, the synchrotron radio emission of W50 and the lobes is not expected to contribute significantly to the cooling of electrons or production of high-energy $\gamma$-rays. In equation~\ref{eqn:transport} we have ignored the diffusion of electrons as it is a slower process than cooling for TeV electrons, and also because doing so saves computing time. Assuming that electrons are injected with a simple power-law spectrum constantly over time, $Q_e(\gamma_e, t)=Q_{e, 0}\,\gamma_e^{-\alpha}$, the solution of equation~\ref{eqn:transport} can be written as
\beq
N(\gamma_e, t) = \frac{Q_{e,0}}{\gamma_e^2}\int_{t_{\rm min}}^t dt_i\,\left[\gamma_e^{-1}-\nu\,(t-t_i)\right]^{\alpha-2}
\eeq
where $t_{\rm min}={\rm max}[0, \, t-\nu^{-1}\,(\gamma_e^{-1} - \gamma_{e, \rm max}^{-1})]$ is the earliest time that an electron with $\gamma_e$ can be injected and still not cooled after time $t$, and $\gamma_{e,\rm max}$ is the maximum electron energy.

\end{document}